\documentclass[aps,prd,twocolumn,groupedaddress,showpacs,preprintnumbers,nofootinbib]{revtex4-1}

\usepackage{graphicx,color,amsmath,amssymb}

\begin{document}

\preprint{ULB-TH/15-25}

\title{Asymmetric Dark Matter in the Sun and Diphoton Excess at the LHC}

\author{P.~S.~Bhupal Dev}
%\email[]{bhupal.dev@mpi-hd.mpg.de}

\affiliation{Physik-Department T30d, Technische Univertit\"{a}t M\"{u}nchen, 
James-Franck-Stra\ss e 1, D-85748 Garching, Germany}
\affiliation{Max-Planck-Institut f\"{u}r Kernphysik, Saupfercheckweg 1, D-69117 Heidelberg, Germany} 

\author{Daniele Teresi}
%\email[]{daniele.teresi@ulb.ac.be}

\affiliation{Service de Physique Th\'eorique, Universit\'e Libre de Bruxelles,\\
Boulevard du Triomphe, CP225, 1050 Brussels, Belgium} 

%\date{December 22, 2015}

\begin{abstract}
It has been recently pointed out that a momentum-dependent coupling of the asymmetric Dark Matter (ADM) with nucleons can explain the broad disagreement between helioseismological observables and the predictions of standard solar models. In this paper, we propose a minimal simplified ADM model consisting of a scalar and a pseudoscalar mediator, in addition to a Dirac fermionic DM, for generating such momentum-dependent interactions. Remarkably, the pseudoscalar with mass around 750 GeV can simultaneously explain the solar anomaly and the recent diphoton excess observed by both ATLAS and CMS experiments in the early $\sqrt s=13$ TeV LHC data. In this framework, the total width of the resonance is naturally large, as suggested by the ATLAS experiment, since the resonance mostly decays to the ADM pair. The model predicts the existence of a new light scalar in the GeV range, interacting with quarks, and observable dijet, monojet and $t\bar{t}$ signatures for the 750 GeV resonance at the LHC.

\end{abstract}


\maketitle

\section{Introduction}
Based on the early $\sqrt s=13$ TeV LHC data, both ATLAS and CMS experiments have reported an excess in the diphoton invariant mass distribution around 750 GeV~\cite{atlas, CMS:2015dxe,atlasmoriond,CMS:2016owr}. Although the significance of this excess is only around $3\: \sigma$ and more statistics is required to draw a firm conclusion, it has already triggered an avalanche of theoretical speculations in terms of a new resonance $X$; see e.g.~\cite{Mambrini:2015wyu, Backovic:2015fnp, Barducci:2015gtd, Chao:2015nsm, Buttazzo:2015txu, Franceschini:2015kwy, Pilaftsis:2015ycr, McDermott:2015sck, Ellis:2015oso, DiChiara:2015vdm,Low:2015qep, Molinaro:2015cwg,Becirevic:2015fmu,Fichet:2015vvy, Nakai:2015ptz,Knapen:2015dap,Harigaya:2015ezk,Falkowski:2015swt, Berthier:2015vbb,Ahmed:2015uqt, Heckman:2015kqk,Chakrabortty:2015hff,Cao:2015twy,Huang:2015evq,Bardhan:2015hcr, Angelescu:2015uiz,Bellazzini:2015nxw,Gupta:2015zzs,Petersson:2015mkr,Dutta:2015wqh, Cho:2015nxy, Han:2015cty}. There are some key points to bear in mind: (i) Since a spin-1 object cannot decay to two photons due to the Landau-Yang theorem, the simplest interpretation of this new resonance is in terms of a spin-0 object with 750 GeV mass.\footnote{See also~\cite{Chu:2012qy,Jaeckel:2012yz} for earlier investigations.}
(ii) The best-fit cross-section values
for $\sigma(pp\to X\to \gamma\gamma)\sim {\cal O}(10~{\rm fb})$~\cite{atlas, CMS:2015dxe} suggest that the new resonance should have a rather large effective coupling to the Standard Model (SM) quarks/gluons. In order to reconcile the diphoton signal with the non-observation of any corresponding signal in the hadronic final states, new physics must be involved in the decay process, in addition to the 750 GeV resonance, while the production can still be due to quark-antiquark annihilation at the tree-level, unless it is Yukawa-suppressed. (iii) In addition, ATLAS has reported a rather large decay width of $\Gamma_X \sim 45$ GeV for the resonance~\cite{atlas} which, if confirmed, means that $X$ must have a sizable partial decay width to experimentally challenging/invisible final states, since the partial decay width to $\gamma\gamma$ is loop-suppressed and can hardly make up for the observed width.  

It is interesting to note that if $X$ couples sizably and decays dominantly to cosmologically stable Dark Matter (DM) particles $\chi$, it can easily have a large decay width~\cite{Mambrini:2015wyu, Backovic:2015fnp}. The existence of a particle DM  is strongly   motivated by several astrophysical and cosmological observations; see e.g.~\cite{Bertone:2004pz} for a review. Although its basic properties such as mass, spin and couplings are still unknown, there are several well-motivated reasons to believe that the DM is sufficiently light so that the tree-level decay of the new resonance $X$ to a pair of DM particles is kinematically allowed, while satisfying all current experimental constraints.  In particular, the fact that the observed DM and baryon abundances in our Universe are quite similar, i.e. $\rho_\chi/\rho_B \approx 5$~\cite{Ade:2015xua}, suggests that these two seemingly disparate quantities might be related in some way. This is naturally realized in asymmetric DM (ADM) scenarios (for reviews, see e.g.~\cite{Davoudiasl:2012uw,Petraki:2013wwa,Zurek:2013wia}), where the asymmetry in the number density of DM  over anti-DM is similar to the baryonic sector, thus pointing towards a light DM with $m_\chi \sim  5m_p \simeq 5$ GeV, where $m_p$ is the proton mass. Furthermore, it was recently shown~\cite{Vincent:2014jia, Vincent:2015gqa} that if the ADM has a momentum-dependent coupling to nucleons,  it can resolve the broad disagreements in solar physics between helioseismological observables and predictions of standard solar models.\footnote{A reassessment of the solar abundance problem in light of a newly determined lower limit on the solar metallicity~\cite{Vagnozzi:2016cmr} claims that the ADM solution is no longer plausible. However, this is by no means a full-proof claim and a number of arguments against it have been presented recently in~\cite{Serenelli:2016nms}. Given that the main focus of our work is on a simplified model which connects the ADM physics to the LHC diphoton excess,  we do not intend to enter into the ongoing debate of solar abundance, rather simply use the ADM solution of~\cite{Vincent:2014jia, Vincent:2015gqa} as our benchmark choice. } 

In this paper, we present a {\it minimal} simplified model for a momentum-dependent ADM, where we introduce just two additional degrees of freedom, viz. a real scalar field and a real pseudoscalar field, which have a small mixing with each other. We argue that, while the mostly scalar mass eigenstate must be at the GeV scale to explain the above-mentioned anomalies in solar physics, the pseudoscalar is required to have a large coupling to the DM, so that it can easily be the 750 GeV resonance observed at the LHC.  We further note that the coupling of the pseudoscalar to SM quarks must be very small to avoid constraints from nuclear Electric Dipole Moment (EDM); therefore, the dominant production of the pseudoscalar field 
at the LHC 
proceeds through its mixing with the scalar field. As discussed above, the decay process into two photons must involve additional new physics to enhance the diphoton branching ratio, as compared to the heavy-quark ones. So far, popular choices include the introduction of exotic vector-like fermions or the presence of a new strongly-interacting sector~\cite{Mambrini:2015wyu, Backovic:2015fnp, Barducci:2015gtd, Chao:2015nsm, Buttazzo:2015txu, Franceschini:2015kwy, Pilaftsis:2015ycr, McDermott:2015sck, Ellis:2015oso, DiChiara:2015vdm,Low:2015qep, Molinaro:2015cwg,Becirevic:2015fmu,Fichet:2015vvy, Nakai:2015ptz,Knapen:2015dap,Harigaya:2015ezk,Falkowski:2015swt, Berthier:2015vbb,Ahmed:2015uqt, Heckman:2015kqk,Chakrabortty:2015hff,Cao:2015twy,Huang:2015evq,Bardhan:2015hcr, Angelescu:2015uiz,Bellazzini:2015nxw,Gupta:2015zzs,Petersson:2015mkr,Dutta:2015wqh, Cho:2015nxy, Han:2015cty}.

\section{The Model}
The minimal simplified model to generate a momentum-dependent DM cross section, via the effective operator $i \: \overline{\chi} \gamma_5 \chi \: \overline{q} q$, consists of a real scalar and a real pseudoscalar  mediator, denoted respectively by $\phi$ and $\phi_P$, which mix with each other. The mass and interaction Lagrangian of the simplified model is
\begin{align}\label{eq:model}
- \mathcal{L} \ &\supset \ \frac{m_\phi^2}{2} \, \phi^2 \; + \; \frac{m_P^2}{2} \, \phi_P^2 \; + \; \mu^2 \, \phi_P \phi \; + \; m_\chi \overline{\chi} \chi \notag\\
&+ \; g_\phi \, \phi \, \overline{q} q \; + \; i\, g_P \, \phi_P \, \overline{q} \gamma_5 q \; + \; i\, h \, \phi_P \, \overline{\chi} \gamma_5 \chi \;,
\end{align}
where, for simplicity, we have taken the Yukawa couplings to the SM quarks as flavour-blind.\footnote{This simplifying assumption is however not crucial for our analysis. All we need are sizable couplings to at least one first and one second-generation quark, for the solar-physics argument and the diphoton explanation, respectively.} In a full $SU(2)$-invariant model, these couplings may originate from effective operators of the form $\frac{1}{\Lambda}\, \bar{Q}_L \, \Phi \, q_R \, \phi$, when the (possibly beyond the SM) scalar doublet $\Phi$ gets a vacuum expectation value below the cut-off scale $\Lambda$. In this case, sizable couplings of $\phi$ to light quarks can be generated by its mixing with the heavy scalar doublet $\Phi$, having Yukawa-type couplings with the quark sector. The vacuum expectation value of the latter must be small, in order to comply with electroweak precision data, but a sizable mixing between $\phi$ and $\Phi$ can be realized anyway, at the price of some fine-tuning of the parameters of this extra scalar sector. 
%these additional fields may originate from suitable $SU(2)_L$ multiplets, but the detailed  model-building aspects are not crucial for our subsequent discussion. 

In the spirit of a simplified-model approach followed here, in \eqref{eq:model} we have included only the couplings relevant for the following discussion. In particular, we assume that any mixed quartic couplings between the SM Higgs and the new spin-0 particles, which are {\it a priori} allowed by the symmetries of the model Lagrangian, are sufficiently small so that they do not affect significantly the SM Higgs phenomenology. Also, for the purposes of this work, we have not included a scalar coupling to DM, because this would generate a momentum-\emph{independent} cross section. However, a small coupling of this kind could be potentially included in a more general analysis, by making sure that the momentum-independent part is sub-dominant.

The scalar and pseudoscalar fields mix into the mass eigenstates $\phi_S$ and $\phi_A$, mostly scalar and pseudoscalar, respectively. The mixing angle $\alpha$, in the limit $m_S \ll m_A$ of interest in the following, is approximately given by
\begin{equation}
\tan \alpha \ \simeq \ \frac{\mu^2}{m_A^2} \;.
\end{equation}
The couplings of the mass eigenstates to quarks and DM are easily found as
\begin{align}
- &\mathcal{L} \ \supset \ g_\phi \, c_\alpha \, \phi_S \, \overline{q} q \; + \;  g_\phi \, s_\alpha \, \phi_A \, \overline{q} q \; - \; i\, g_P \,  s_\alpha \, \phi_S \, \overline{q} \gamma_5 q  \notag\\
&+ \; i\, g_P \,  c_\alpha \, \phi_A \, \overline{q} \gamma_5 q \; - \; i\, h \,  s_\alpha \, \phi_S \, \overline{\chi} \gamma_5 \chi \; + \; i\, h \,  c_\alpha \, \phi_A \, \overline{\chi} \gamma_5 \chi \;,
\end{align}
where we have introduced the abbreviations $s_\alpha \equiv \sin \alpha$ and $c_\alpha \equiv \cos \alpha$.

\section{ADM from solar physics}
The basic idea is that collisions between ADM and nuclei can lead to capture and accumulation of DM in large quantities in the solar core, if the collisions result in sufficient energy transfer to bring down the DM velocity below the local escape velocity~\cite{Vincent:2014jia, Vincent:2015gqa, Vagnozzi:2016cmr, Serenelli:2016nms, Frandsen:2010yj, Cumberbatch:2010hh, Taoso:2010tg, Lopes:2014aoa, Blennow:2015xha}. Optimal energy transfer, and hence, optimal capture rate, occurs for DM masses close to that of the solar composition, which is mostly hydrogen and helium. It is interesting that this is roughly the same mass range expected in generic ADM models to explain the 5:1 relic DM-to-baryon density~\cite{Davoudiasl:2012uw,Petraki:2013wwa,Zurek:2013wia}. Momentum-dependent DM-nucleon scattering in the Sun leads to an additional efficient mechanism, alongside photons, for heat transport from the solar core to the outer regions, thereby affecting various helioseismological observables. 

In Refs.~\cite{Vincent:2014jia,Vincent:2015gqa}, a formally $6\sigma$ preference for momentum-dependent ADM was presented, which resolved the long-standing anomalies of the standard solar models in describing the observed sound-speed profile, convective zone depth, surface helium abundance and small frequency separations in the Sun. In particular, they found the best-fit values of the DM mass and interaction cross section with nucleons respectively as 
\begin{equation}\label{eq:bestfit}
m_\chi \ = \ 3\,\textrm{GeV} \;, \quad \sigma_{\rm DD} \ = \ (|\vec{q}|/ 40 \, \textrm{MeV})^2 \, 10^{-37} \mathrm{cm}^2 \;,
\end{equation}
where $\vec{q}$ is the 3-momentum exchanged in the ``direct-detection'' process $\chi N \to \chi N$, i.e. the non-relativistic scattering of the DM with the nucleon $N$. Although this best-fit point~\eqref{eq:bestfit} has been ruled out very recently using the latest CRESST-II data~\cite{Angloher:2016jsl}, there still remains a sizeable part of the momentum-dependent ADM parameter space~\cite{Vincent:2015gqa} which provides significant improvement with respect to standard solar models. For concreteness, we choose a benchmark value of  $
m_\chi = 2\,\textrm{GeV}$,\footnote{A 2-3 GeV DM is prone to evaporation from the solar interior~\cite{Gould:1987ju, Busoni:2013kaa, Kouvaris:2015nsa}. The evaporation rate depends on the
interaction cross-section, mean free path and thermal regime. In the analysis of~\cite{Vincent:2015gqa}, it is argued that this is small for nuclear scattering cross-sections allowed by direct detection. A full kinematic analysis would be necessary to assess this issue more quantitatively, but this is beyond the scope of this work.} 
 while keeping $\sigma_{\rm DD}  =  (|\vec{q}|/ 40 \, \textrm{MeV})^2 \, 10^{-37} \mathrm{cm}^2$, as in Eq.~\eqref{eq:bestfit}, which is well within the preferred ADM parameter space~\cite{Vincent:2015gqa}, as well as consistent with the CRESST-II limits~\cite{Angloher:2016jsl}.  According to our estimate, this benchmark point is also consistent with the latest CDMSlite results~\cite{Agnese:2015nto}, after extrapolating their limit on the momentum-independent cross section to the momentum-dependent cross section considered here, by a simple rescaling of the typical 3-momentum exchanged.

At the quark level, $\sigma_{\rm DD}$ in our model is given by the process $\chi q \to \chi q$, with $t-$channel exchange of the mediators $\phi_{S}$, $\phi_A$. In particular, for $q^2 \ll m^2_{S,A}$, the following local term is generated  in the quark-level effective Lagrangian:
\begin{equation}
\frac{\sin 2 \alpha}{2} \, h \, g_\phi \left( \frac{1}{m_S^2} - \frac{1}{m_A^2}\right) i \, \overline{\chi} \gamma_5 \chi \, \overline{q} q \; \equiv \; J \, i \, \overline{\chi} \gamma_5 \chi \, \overline{q} q \;.
\end{equation}
We perform the matching to the nucleon effective Lagrangian in the standard way. The effective coupling to the proton is found to be
%\begin{equation}
$0.47 \, J \, i \, \overline{\chi} \gamma_5 \chi \, \overline{N} N$,
%\end{equation}
and for the non-relativistic differential cross section we obtain
\begin{equation}
\frac{d \sigma}{d \cos \theta} \ = \ \frac{1}{8 \pi} \, \frac{m^2_N}{(m_\chi + m_N)^2} \, (0.47 \, J \, |\vec{q}|)^{\,2} \;.
\end{equation}
For the benchmark values chosen above, we thus obtain the constraining relation for the model parameters: 
\begin{equation}\label{eq:constraint}
\frac{\sin 2 \alpha}{2} \, h \, g_\phi \left| \frac{\mathrm{GeV}^2}{m_S^2} - \frac{\mathrm{GeV}^2}{m_A^2}\right| \ \simeq \ 9.6 \times 10^{-3} \;. 
\end{equation}
This equation provides the constraint on the model parameters needed to account for the benchmark value above. Thus, in order to reproduce the results of \cite{Vincent:2015gqa}, we need relatively large mixing and couplings of $O(10^{-3}\-10^{-2})$, and at least one of the two mediators in the GeV range. In the following, we will use \eqref{eq:constraint} to fix the value of the scalar coupling to the quarks $g_\phi$ in terms of the other model parameters. 

\section{EDM constraints}

\begin{figure}[t!]
\includegraphics[width=4.5cm]{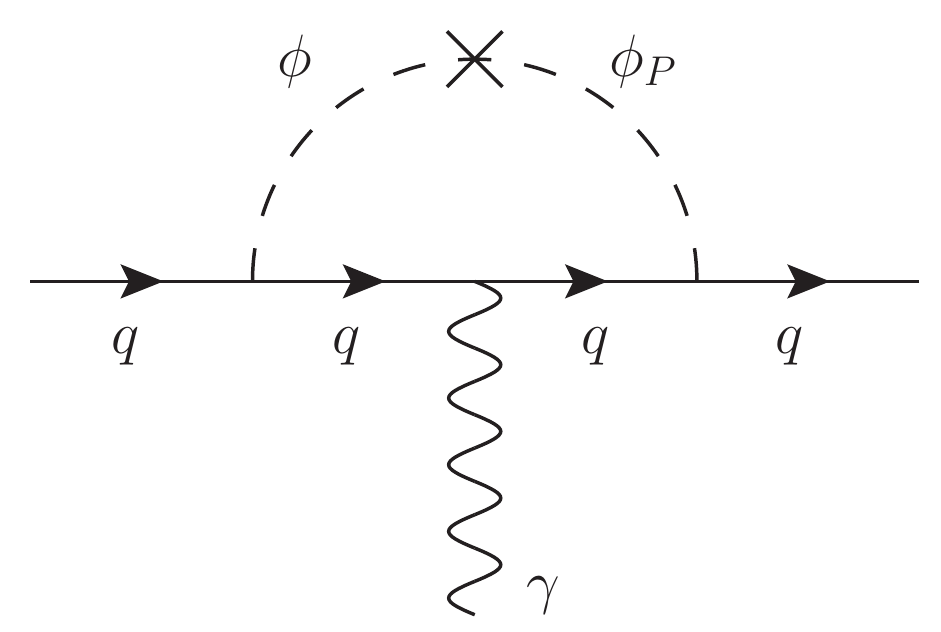}
\caption{1-loop diagram contributing to the neutron EDM. \label{fig:EDM}}
\end{figure}

The mixing of the scalar and pseudoscalar bosons induces an EDM for the quarks, by means of the 1-loop diagram in Figure \ref{fig:EDM}. For the EDM of the down quark, we find
\begin{align}
d_d \ &\simeq \ 8.3 \,\times\, 10^{-19} \, e \, \mathrm{cm} \,\times  \left(4.55 + \ln \frac{\widetilde{m}}{\rm GeV}\right) \notag \\
& \times \, \frac{\sin 2 \alpha}{2} \,  g_P \, g_\phi \left( \frac{\rm GeV^2}{m_S^2} - \frac{\rm GeV^2}{m_A^2}\right) ,
\end{align}
where $\widetilde{m}$ is the mass scale of the lightest mediator. The EDM of the neutron can be related to the one of the quarks as given, e.g., in Ref.~\cite{Hisano:2012sc}. Here, for a rough estimate, we approximate it as $d_n \approx 0.5 \, d_d$. By requiring $|d_n| < 3 \times 10^{-26} \, e\,\mathrm{cm}$ at 90\% CL~\cite{Afach:2015sja}, and combining with \eqref{eq:constraint}, we obtain the EDM bound
\begin{equation}
\frac{g_P}{g_\phi} \ \lesssim \ 10^{-6} \;.
\end{equation}
Thus, the pseudoscalar coupling to quarks $g_P$ will play no role in what follows, and we may set it to zero to simplify the model phenomenology. Also, we assume negligible couplings to the SM leptons, in order to avoid a plethora of low-energy constraints in the lepton sector, most notably from the electron EDM~\cite{Baron:2013eja}. 

\section{CMB constraints}
\begin{figure}[t!]
\includegraphics[width=8cm]{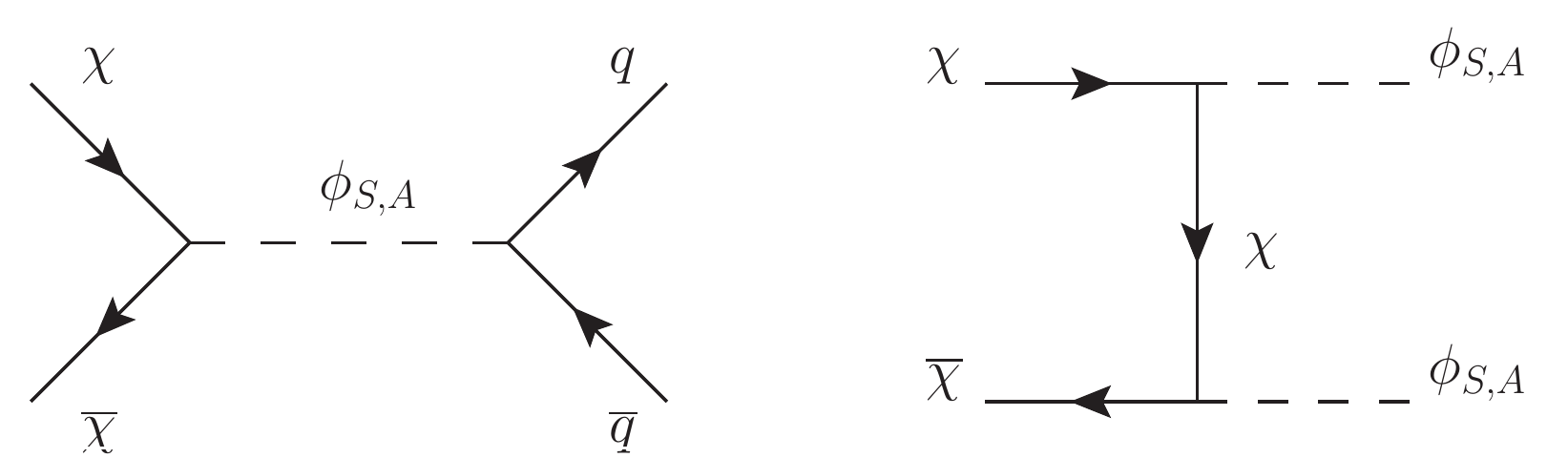}
\caption{Annihilation processes relevant for the depletion of the thermal symmetric component of the DM.\label{fig:annihilation}}
\end{figure}
For ADM it is not possible to impose directly the successful prediction of the observed relic density, unless one considers an explicit mechanism for the generation of the DM asymmetry~\cite{Davoudiasl:2012uw,Petraki:2013wwa,Zurek:2013wia}. This is because the relic density is typically proportional to the primordial asymmetry, assuming that annihilation processes dominantly conserve the DM minus anti-DM number. However, the successful annihilation of the symmetric thermal component allows one to put a \emph{lower} bound on the relevant annihilation cross sections \cite{Lin:2011gj}. As shown in Figure \ref{fig:annihilation}, for the model under consideration, the dominant annihilation processes are given by the $s-$channel annihilation into quarks and $t-$ and $u-$channel annihilations into a pair of mediators, the latter if kinematically allowed. We will denote their cross sections by $\sigma_{qq}$ and $\sigma_{\phi \phi}$, respectively.

A too large symmetric component of DM at the epoch of recombination would cause a sizable effect on the CMB, coming from DM-anti-DM annihilation. Assuming an ionization efficiency factor $f=1$, in Ref.~\cite{Lin:2011gj} this ``indirect detection'' bound is found to be
\begin{equation}\label{eq:CMB}
\langle \sigma v \rangle_{\rm CMB} \ < \ (1.2 \, \times \, 10^{-27} \, {\rm cm}^3\, {\rm s}^{-1}) \, \frac{m_\chi}{\rm GeV} \, \frac{1}{r_\infty} \;,
\end{equation}
where $\langle \sigma v \rangle_{\rm CMB}$ is the cross section times relative velocity, thermally averaged at the epoch of recombination, and $r_\infty$ is the ratio of anti-DM to DM energy density.

On the other hand, in order to fit the observed DM relic density, the annihilation cross section at the freezeout epoch needs to be, approximately \cite{Graesser:2011wi}
\begin{equation}\label{eq:relic}
\langle \sigma v \rangle_{\rm f} \ \simeq \ (5 \, \times \, 10^{-26} \, {\rm cm}^3{\rm s}^{-1}) \,  \ln \frac{1}{r_\infty} \;.
\end{equation}
The annihilation into quarks is an $s$-wave process, and thus $\langle \sigma_{qq} v \rangle_{\rm f} \simeq \langle \sigma_{qq} v \rangle_{\rm CMB}$. Instead, the annihilation into the scalar mediators is $p$-wave suppressed, being proportional to $v^2$, and hence, $\langle \sigma_{\phi \phi} v \rangle_{\rm CMB} \approx 10^{-15} \langle \sigma_{\phi \phi} v \rangle_{\rm f}$, having taken $v_{\rm} \approx 0.3$ and $v_{\rm CMB} \approx 10^{-8}$. Therefore, eliminating the anti-DM-to-DM ratio $r_\infty$ from \eqref{eq:CMB} and \eqref{eq:relic}, and imposing $m_\chi = 2 \, \mathrm{GeV}$ for our benchmark value, we find the CMB bound 
\begin{equation}\label{eq:CMBbound}
\langle \sigma_{qq} v \rangle_{\rm f} \ < \ (2.4 \, \times \, 10^{-27} \, {\rm cm}^3{\rm s}^{-1}) \,  \exp \left[\frac{\langle \sigma_{qq} v \rangle_{\rm f} + \langle \sigma_{\phi \phi} v \rangle_{\rm f}}{5 \, \times \, 10^{-26} \, {\rm cm}^3{\rm s}^{-1}}\right],
\end{equation}
which will be used in the next section. Notice that the bound occurs far from the $s$-channel resonance for $\sigma_{qq}$. At the resonance the bound is more easily satisfied, since $\langle \sigma_{qq} v \rangle_{\rm f}$ appears in the exponent on the right-hand side of~\eqref{eq:CMBbound}.

\section{Diphoton excess}
In order to fit the diphoton excess, we fix $m_A=750$ GeV and $\Gamma_A \simeq 45$ GeV, as suggested by the ATLAS result~\cite{atlas}. The best-fit signal cross sections are given by 
\begin{align}
\sigma(pp\to X\to \gamma\gamma) & =\left\{\begin{array}{cc} 
(10\pm 3)~{\rm fb} & ({\rm ATLAS}) \\
 (6\pm 3)~{\rm fb} & ({\rm CMS})
\end{array}\right.,
\label{sig}
\end{align}
Thus, for our numerical purpose, we take the conservative range of $(8\pm 5)$ fb.  The dominant production channels in our case are the quark-antiquark annihilations to the pseudoscalar $\phi_A$, induced by the scalar-pseudoscalar mixing angle $\alpha$. At $\sqrt s=13$ TeV, we obtain the leading-order cross section $\sigma(pp\to \phi_A)=(433~{\rm pb}) \,g^2_\phi s^2_\alpha$, obtained using {\tt MadGraph5}~\cite{Alwall:2014hca} with {\tt NNPDF2.3} parton distribution functions~\cite{Ball:2012cx}. Note that the gluon-gluon fusion or any other loop-induced new physics contribution will be sub-dominant, as compared to the tree-level $q\bar{q}$ annihilation, unless we introduce a large number of new colored objects to run in the loop. 

As for the decay process $\phi_A\to \gamma\gamma$, some new physics must be present, in addition to the simplified model considered here, to enhance its partial decay width $\Gamma_{\gamma \gamma}$ to the level that $\sigma(pp\to \phi_A)\times {\rm BR}_{\gamma\gamma}$ is within the observed range. Popular choices, considered so far, include vector-like fermions and/or strong dynamics~\cite{Mambrini:2015wyu, Backovic:2015fnp, Barducci:2015gtd, Chao:2015nsm, Buttazzo:2015txu, Franceschini:2015kwy, Pilaftsis:2015ycr, McDermott:2015sck, Ellis:2015oso, DiChiara:2015vdm,Low:2015qep, Molinaro:2015cwg,Becirevic:2015fmu,Fichet:2015vvy, Nakai:2015ptz,Knapen:2015dap,Harigaya:2015ezk,Falkowski:2015swt, Berthier:2015vbb,Ahmed:2015uqt, Heckman:2015kqk,Chakrabortty:2015hff,Cao:2015twy,Huang:2015evq,Bardhan:2015hcr, Angelescu:2015uiz,Bellazzini:2015nxw,Gupta:2015zzs,Petersson:2015mkr,Dutta:2015wqh, Cho:2015nxy, Han:2015cty}. Here, we do not wish to reiterate these interpretations, and instead proceed in a model-independent way, by presenting our results in terms of the value of $\Gamma_{\gamma \gamma}$.
The other relevant decay channels are $\phi_A\to q\bar{q}$ (with $q=u,d,c,s,b,t$) and $\phi_A\to \chi \bar{\chi}$, with the following tree-level decay rates:  
\begin{align}
\Gamma_{q\bar{q}} \ &= \ \frac{N_c \, g_\phi^2\,s^2_\alpha}{8\pi} \,m_A \left(1-\frac{4m_q^2}{m_A^2}\right)^{3/2}, \label {eq:jj}\\
\Gamma_{\chi\bar\chi} \ &= \ \frac{h^2 \, c_\alpha^2}{8\pi} \, m_A\left(1-\frac{4m_\chi^2}{m_A^2}\right)^{3/2}.
\end{align}
We have ignored sub-dominant decay modes such as $\phi_A\to \gamma Z, \, ZZ, \, gg$.\footnote{Although $\Gamma_{gg} > \Gamma_{\gamma \gamma}$, the contribution of the former to the experimental signals is sub-leading, since the dijet rate is dominated by the tree-level decay into quarks.} Thus, the total decay width is simply given by 
\begin{align}
\Gamma_A & \ \simeq \ \Gamma_{\gamma\gamma}+\Gamma_{q\bar{q}}+\Gamma_{\chi\bar{\chi}} \ \simeq \  45~{\rm GeV} \;.
\label{tot}
\end{align}
We fit the signal cross section for different values of $\Gamma_{\gamma \gamma}$.
The  95\% CL upper limits from $\sqrt s=8$ TeV LHC data on dijet~\cite{CMS:2015neg, Aad:2014aqa} and $t\bar{t}$~\cite{Khachatryan:2015sma} signal cross sections of 2.5 pb and 450 fb put upper limits on  $g_{\phi}s_\alpha \lesssim 0.20 $ and $\lesssim 0.22 $, respectively. These are obtained from the leading-order cross section $\sigma(pp\to \phi_A)=(159~{\rm pb}) \,g^2_\phi s^2_\alpha$ at $\sqrt s=8$ TeV and using the branching ratios as given by~\eqref{eq:jj}. 
%Taking into account these limits, we obtain the absolute lower bound $\Gamma_{\gamma \gamma} \gtrsim 8 \, \mathrm{MeV}$ to explain the diphoton excess, while being consistent with the dijet constraints. 

The 95\% CL upper limit on the $\gamma\gamma$ cross section of 1.5 fb from the $\sqrt s=8$ TeV LHC searches~\cite{CMS:2014onr, Aad:2015mna} implies an upper limit of $g_\phi s_\alpha\lesssim 0.02/\sqrt{\Gamma_{\gamma\gamma}/{\rm GeV}}$ for a flavour-blind coupling. Even though this bound seems to be very stringent, there is still some allowed region of parameter space to explain the observed diphoton signal at $\sqrt s=13$ TeV. Here, we stress that the compatibility between the diphoton results from the $\sqrt s=8$ and 13 TeV data can be enhanced depending on the detailed flavour structure of the coupling $g_\phi$, and in particular, for a larger coupling to $c,s,b$-quarks, for which the production cross section ratio at $\sqrt s=13$ and 8 TeV LHC is larger~\cite{Franceschini:2015kwy, Gupta:2015zzs}.  

For a large banching ratio of the resonance decay to DM, as required here to explain the large total width, the constraints from DM searches at the LHC via monojets~\cite{Khachatryan:2014rra, Aad:2015zva} turn out to be significant too~\cite{Falkowski:2015swt, Barducci:2015gtd}. For instance, using the most stringent 95\% CL upper limit on the monojet cross section of 3.4 fb~\cite{Aad:2015zva}, we obtain $g_\phi s_\alpha\lesssim 0.04/\sqrt{{\rm BR}(\phi_A\to \chi\bar{\chi})}$.\footnote{The corresponding limits on the spin-independent DM-nucleon cross section derived in~\cite{Aad:2015zva} cannot be applied directly to our case, since they did not consider the momentum-dependent operator. Also, one might wonder whether the $pp\to \phi_S\to \chi\bar\chi$ process could lead to more stringent monojet constraints. However, for the selection cuts used in the monojet searches~\cite{Khachatryan:2014rra, Aad:2015zva}, the possibility of a resonant contribution from an on-shell $\phi_S$ can be readily ruled out, and the ensuing limits from an off-shell $\phi_S$ will be much weaker compared to those derived using the resonant $\phi_A$ production mentioned above.}

\begin{figure}[t]
\includegraphics[width=0.4\textwidth]{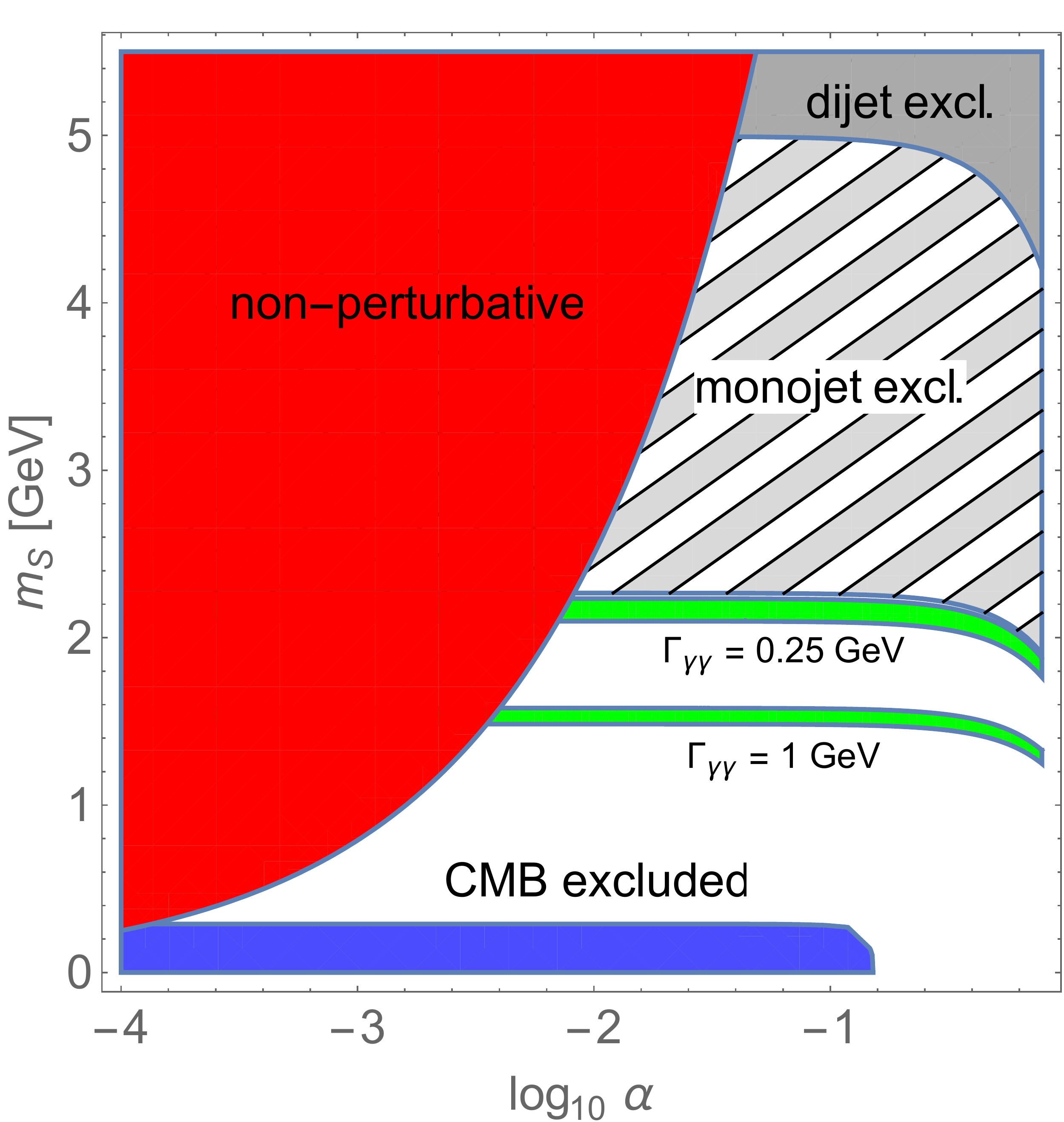}
\caption{The parameter space of our ADM model, constrained from perturbativity and CMB arguments, in addition to the non-observation of monojet, dijet and diphoton signals at the $\sqrt s=8$ TeV LHC. We exhibit the allowed parameter space giving the observed diphoton signal (green shaded regions), for two benchmark values of  $\Gamma_{\gamma \gamma} = 0.25 $ and $1 \, \mathrm{GeV}$. 
\label{fig:combined}}
\end{figure}

In order to translate the implications of the diphoton excess in this model to the parameter space relevant for ADM, we fix the $\phi_A\chi\bar{\chi}$ coupling from the total decay-width formula \eqref{tot}, obtaining $h \simeq 1.2$, which is well within the perturbative limit. Plugging this into~\eqref{eq:constraint}, we show the allowed parameter space in the scalar mixing--mass plane satisfying all experimental constraints in Figure~\ref{fig:combined}. Here, the perturbativity condition $g_\phi < 5$ excludes the red shaded region. The blue and gray shaded regions are excluded from CMB and dijet constraints, respectively, as discussed above. The gray-meshed region represents the monojet exclusion region from the $\sqrt s=8$ TeV data, assuming ${\rm BR}(\phi_A\to \chi\bar{\chi})$ close to unity and a conservative estimate of the signal acceptance times efficiency of 80\%. For the presentation of our results, we have chosen the benchmark values $\Gamma_{\gamma \gamma} = 0.25$ and $1 \, \mathrm{GeV}$,  for which the diphoton favoured regions are shown by the upper and lower green shaded regions, respectively, after having imposed the 8 TeV diphoton constraint discussed above. 

Here, we would like to mention that if the large width requirement $\Gamma_A \simeq 45$ GeV is lifted, the monojet constraint become less stringent. In this case, we find an absolute {\it lower} bound $\Gamma_{\gamma \gamma} > 0.5$ MeV, coming from the dijet and CMB constraints.

\begin{figure}[t]
\includegraphics[width=0.42\textwidth]{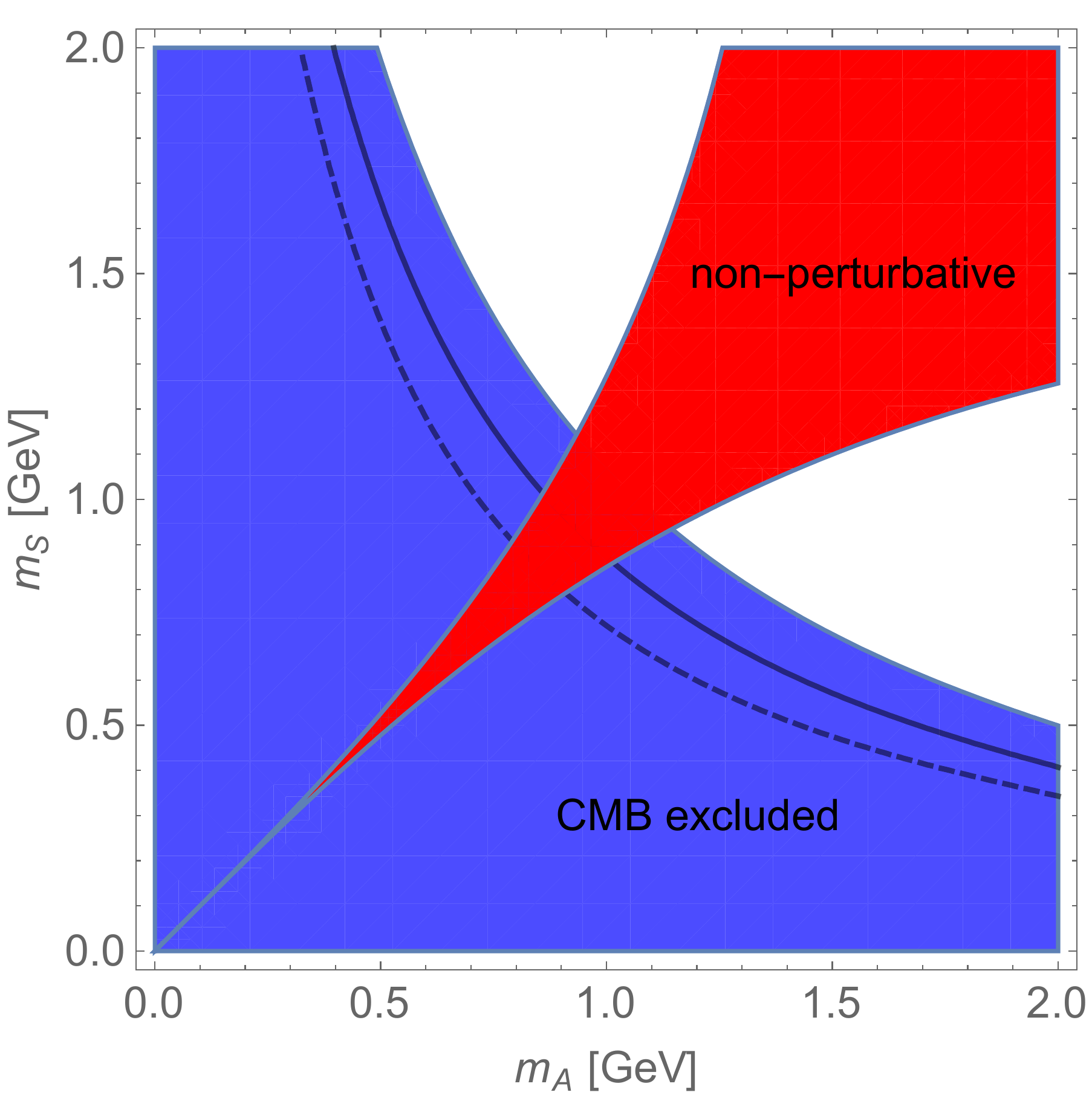}
\caption{The parameter space of our ADM model, constrained from perturbativity and CMB observations, {\em without} identifying the pseudoscalar with the  $750 \, \mathrm{GeV}$ resonance. We have chosen the remaining parameters as $h = 0.1$, $\alpha = 0.05$, for illustration. The black dashed and solid lines denote the anti-DM-to-DM energy density ratio $r_\infty$ of 10\% and 1\%, respectively. 
\label{fig:light_mediators}}
\end{figure}

Before concluding, we should also remark that in case the diphoton signal goes away with more statistics, the mostly pseudoscalar state in our model does not necessarily have to be this heavy. Just as an illustration, we show in Figure~\ref{fig:light_mediators} the allowed parameter space when both the scalar and pseudoscalar mediators are light for a typical choice of the coupling $h=0.1$ and the scalar-pseudoscalar mixing $\alpha=0.05$. As expected, the CMB constraints are more stringent in this case, but there is still a large parameter space that is allowed.

\section{Conclusion}
We have presented a minimal simplified model of ADM with momentum-dependent interaction with nucleons, which resolves some pronounced discrepancies in solar physics. At the same time, we can also interpret the recent diphoton excess at the LHC as due to the resonant production of one of the scalar mass eigenstates in this model, which dominantly decays into the asymmetric DM pair to give a broad resonance. As discussed extensively in the literature, the decay process into a pair of photons must involve additional new physics, on top of this simplified model. In this respect, we have chosen to proceed in a model-independent way, by parametrizing our results in terms of the diphoton partial decay width and have shown in Figure~\ref{fig:combined} the preferred parameter space, satisfying the relevant experimental constraints from monojet and dijet searches at the LHC, as well as the CMB and perturbativity constraints. 

The model predicts the existence of a new scalar in the GeV range, interacting with quarks. The existence of such light scalars is still allowed by low-energy and fixed-target experiments, and could be potentially tested at the proposed SHiP  experiment~\cite{Alekhin:2015byh} or in future B-factories~\cite{barlow}. 
Moreover, if the 750 GeV resonance persists in the future analyses with more statistics, the model predictions for observable monojet, $t\bar{t}$ and dijet signals can be used to test this ADM hypothesis at the LHC. This will be complementary to the future direct detection prospects of a momentum-dependent DM interaction with nuclei.

\section{Acknowledgements} 
We thank Michel Tytgat, Jean-Marie Fr\`ere and Aaron Vincent for useful discussions. This work of P.S.B.D. was supported in part by a TUM University Foundation Fellowship, as well as by the DFG with grant RO 2516/5-1. The work of D.T. is funded by the Belgian Federal Science Policy through the Interuniversity Attraction Pole P7/37.

\end{document}